\documentstyle[aps]{revtex}

\begin{document}
\title{{\bf A quantum secret sharing scheme} {\bf among three parties ultilizing
four-qubit Smolin bound entangled state}}
\author{Yafei Yu$^{\thanks{%
corresponding author, e-mail: yfyuks@hotmail.com}a,b}$, Yi Xu$^{a}$and Jin
Liu$^{a}$}
\address{$^{a}$School for Information and Optoelectronic Science and Engineering,\\
South China Normal University,\\
GuangZhou,510006, PR China\\
$^{b}$Laboratory of Photonic Information Technology, South China Normal\\
University, GuangZhou,510006, PR China}
\maketitle

\begin{abstract}
Four-qubit Smolin bound entangled state\cite{Smolin} has a distinct feature:
the state is not distillable when every qubit is seperated from each other;
but it makes two separated qubit entangled if the other qubits group
together. Here the feature is applied to quantum secret sharing, a QSS
protocol similar to Ekert 91 protocol of QKD is proposed. The security
problem, disadvantage and advantageof this protocol are disscused.
\end{abstract}

\subsection{Introduction}

In the past years, with the development of quantum key distribution\cite
{Charlie,Ekert}, quantum secret sharing (QSS)\cite{Hillery} attracts much
attention in both the theoretical and experimental aspects of quantum
communication. QSS is a protocol to split a message to several parts so that
no subset of parts is sufficient to read the message but the entire set is.
For example, suppose that Alice wants to send a secret to two distant
parties, Bob and Charlie. One of them, Bob or Charlie, is not entirely
trusted by Alice, but she knows if they who can come together and carry it
out for her, but she doesn'ttt know whether they are honest or whether there
is an eavesdropper in the channel. She cannot simply send a message to both
by a classical channel, because if the two of them coexist, the honest one
will keep the dishonest one from doing any damage. Instead of giving entire
secret messages to either of them, it may be desirable for Alice to split
the secret messages into two encrypted parts and send each one a part so
that neither individual is able to obtain all of the original information
unless they collaborate. To achieve this end, classical cryptography can use
a technique called secret sharing\cite{Schneier}. QSS is the generation of
this concept to quantum scenario.

Up to now, there are many kinds quantum secret sharing protocols with and
without entanglement. The first QSS\ protocol has been proposed by Hillery
et al\cite{Hillery}, in which three-qubit GHZ (Greenberger-Horne-Zeilinger)
entangled states is employed to allow information splitting and eavesdropper
protection simultaneously. Moreover, Koashi and Imoto considered the
correlation of the two-qubit Bell state in their quantum secret sharing
scheme\cite{Koashi}. Then Bagherinezhad and Karimipour introduced a work for
quantum secret sharing which utilizes the reusable GHZ states as secure
carriers\cite{Bagherinezhad}. Entanglement swapping is another method used
to realize QSS. Karimipour et al. \cite{Karimipour} proposed d-level secret
sharing via entanglement swapping between a generalized cat states for
d-level systems and a generalized Bell states. Other improved versions of
QSS based on entanglement swapping were also presented\cite{Cabello,Juhui
Lee,Zhan-jun Zhang}. Product states are alternative resources for realizing
QSS. Li-Yi Hsu et al \cite{product state} suggested QSS schemes with a
particular set of orthogonal product states in which an unknown quantum
state cannot be determined only by LOCC(local operation and classical
communication) if the order of the local measurements is private. A
BB84-like QSS Scheme was given by the Group Guang-can Guo\cite{Guo-Ping Guo}%
, which security is based on the quantum no-cloning theory. Recently, the
experimental demonstrations of QSS by GHZ states\cite{Yu-Ao Chen} and by a
single qubit\cite{Schmid} were reported, respectively.

In this work, we use a bound entangled state as quantum resource to
accomplish QSS task. Bound entanglement (BE)\cite{horodecki} is a kind of
entanglement in multi-parties system which cannot be distilled to pure
entangled form only by local operations and classical communication (LOCC).
But for some bound entangled states, collective operation between some
subparties induces distillable entanglement shared between the others.
Four-party Smolin entangled state \cite{Smolin} is such state: when two
parties come together, they can by LOCC enable the other two parties to have
some pure entanglement. The Smolin state can be expressed as: 
\begin{eqnarray}
\rho &=&\frac{1}{4}(\left| \Phi ^{+}\right\rangle _{12}\left\langle \Phi
^{+}\right| \otimes \left| \Phi ^{+}\right\rangle _{34}\left\langle \Phi
^{+}\right| +\left| \Phi ^{-}\right\rangle _{12}\left\langle \Phi
^{-}\right| \otimes \left| \Phi ^{-}\right\rangle _{34}\left\langle \Phi
^{-}\right|  \nonumber \\
&&+\left| \Psi ^{+}\right\rangle _{12}\left\langle \Psi ^{+}\right| \otimes
\left| \Psi ^{+}\right\rangle _{34}\left\langle \Psi ^{+}\right| +\left|
\Psi ^{-}\right\rangle _{12}\left\langle \Psi ^{-}\right| \otimes \left|
\Psi ^{-}\right\rangle _{34}\left\langle \Psi ^{-}\right| )
\end{eqnarray}
where we use the usual notation\ for the maximally entangled states of two
qubit.(Bell state) 
\begin{equation}
\left| \Phi ^{\pm }\right\rangle =\frac{1}{\sqrt{2}}(\left| 00\right\rangle
\pm \left| 11\right\rangle )
\end{equation}
\begin{equation}
\left| \Psi ^{\pm }\right\rangle =\frac{1}{\sqrt{2}}(\left| 01\right\rangle
\pm \left| 10\right\rangle )
\end{equation}

In other word, if 1 and 2 particle come together and do the nonlocal Bell
measurement on their systems, they can determine reliably which Bell state
they have since the four Bell states are orthogonal, and 3, 4 have the same
one with 1,2. The feature of Smolin state is helpful for QSS scheme\cite
{Smolin,Augusiak,Augusiak2}.

Our work is organized as follows. In Sec. II we present a QSS scheme
utilizing four party bound entangled states (Smolin states for short), and
analyze the security in the sense of the violation to Bell inequality. In
Sec.III the scheme is compared with others QSS protocol, and the conclusions
are given.

\subsection{Quantum Secret Sharing With Smolin States}

To achieve the purpose of sharing secret, the scheme is divided into three
stages. The first stage is for preparation and distribution of Smolin
states. Alice produces a series of Smolin states and sends two qubit in
every event to Bob and Charlie. Secondly, for checking the security of
quantum channel, Alice chooses randomly some Smolin states from the series.
The three man determine whether the correlation in the Smolin states violate
the Bell inequality or not. Then if the channel is secure, after Alice
operates qubits in hand, Bob and Charlie share Alices' private classical
information which can be revealed only by Bob and Charlie's collaboration.

In the first phase, Alice prepares N copies of four-qubit bound entangled
Smolin states each of which has a corresponding record number. For each
Smolin state qubits 1, 2 are in the possession of Alice, and qubits 3, 4 are
send to Bob and Charlie, respectively. Once Bob and Charlie receive one
qubit, they publicly announce the facts. As a result, Alice has qubits 1, 2
of each Smolin state, Bob has qubit 3 and Charlie qubit 4. And the shared
Smolin states between the three parties will serve as carriers of private
classical information.

In the phase of examining the security of quantum channel, Alice chooses
randomly M copies from the Smolin states, and inform Bob and Charlie which
their record numbers are. Then Alice projects qubit 1 into the basis $\frac{%
\left| x\right\rangle \pm \left| y\right\rangle }{\sqrt{2}}$ and qubit 2
into $\left| x\right\rangle ,\left| y\right\rangle $. Bob and Charlie
measure their qubits 3, 4 in the base of $\left| x\right\rangle ,\left|
y\right\rangle $ and send their results back to Alice, respectively.
Collecting all results of measurements on four qubits, Alice is able to
calculate the value of correlation function $E$ of four qubits in Smolin
state.

For four qubits, a two-setting Bell-inequality similar to standard CHSH \cite
{CHCS} inequality for two particles is given as \cite{Augusiak}:

\begin{equation}
\left| E(1,1,1,1)+E(1,1,1,2)+E(2,2,2,1)-E(2,2,2,2)\right| \leq 2.
\end{equation}

But for four-qubit Smolin entangled state, the correlation function $E_{QM}$
satisfies 
\begin{equation}
\left|
E_{QM}(1,1,1,1)+E_{QM}(1,1,1,2)+E_{QM}(2,2,2,1)-E_{QM}(2,2,2,2)\right| =2%
\sqrt{2}.
\end{equation}
which gives violation, and it is proved that the above violation to
two-setting Bell inequality is maximal \cite{Augusiak}. Therefore, if the
qubits are not directly or indirectly disturbed, Alice gets the Equation
(5), and the quantum channel is secure and can be used for private
communication.

In the phase of transferring information, Alice first measures the qubits
1,2 of the rest (N-M) copies of Smolin states in Bell basis $\left\{ \left|
\Phi ^{\pm }\right\rangle ,\left| \Psi ^{\pm }\right\rangle \right\} $, and
encode the results obtained, for example as $\left| \Phi ^{+}\right\rangle
=00,\left| \Phi ^{+}\right\rangle =01,\left| \Psi ^{+}\right\rangle
=10,\left| \Psi ^{-}\right\rangle =11.$ Hence, after the measurement, Alice
creates a random and private bit string which, simultaneously, is send to
and shared between Bob and Charlie when finishing the measurement is
announced.. To reveal the bit string, Bob and Charlie have to come together
and determine that in which one of four Bell states their qubits are, then
read out Alice's secure bit-string information. Actually and very
importantly, in this phase, the key is created, send and shared just at the
same time. As a result, the task of QSS is achieved.

Then let's go back to consider how to detect possible eavesdropping
attacks.. If there is Eve in the channel who wants to extract out Alice's
private information. As discussed in Ref.\cite{Ekert91}, Eve cannot elicit
any information from the qubits while in transit from Alice to Bob and
Charlie, because the private information is created only after Alice's
measurements on qubits 1,2 and announcement for finishing it. So Eve has to
intercept and clone the qubits 3,4 send to Bob and Charlie, and distribute
one copy between them. However the intervention of Eve will introduce noise
to the original Smolin state. Now the modified Smolin state is expressed in
a general form:

\begin{equation}
\rho ^{noisy}=\frac{1-p}{16}I^{\otimes 4}+p\rho
\end{equation}

where I stands for identity on one-qubit space, $p$ is scaling parameter
which $p\leq \frac{2}{3}$ because of the intervention of Eve. The
corresponding correlation function $E_{QM}$ is amended as follows:

\begin{equation}
\begin{array}{c}
E_{QM}(1,1,1,1)(\rho ^{noisy}(p))+E_{QM}(1,1,1,2)(\rho ^{noisy}(p)) \\ 
+E_{QM}(2,2,2,1)(\rho ^{noisy}(p))-E_{QM}(2,2,2,2)(\rho ^{noisy}(p))=2\sqrt{2%
}p
\end{array}
\end{equation}

For $p\leq \frac{2}{3}$, the value of the Eq.[7] do not violate two-setting
Bell-inequality of four qubits. Therefore, the existence of Eve will be
detected while Alice tests the violation of the correlation function $E$ to
Bell-inequality.

There may be another method for Eve to choose. Eve makes Bell measurement on
qubits 3,4, prepares two copies same as the result, and dispenses one to Bob
and Charlie. It makes that both of Eve and Alice have the same Bell states
as shared between Bob and Charlie. Eve's trick can also be detected if Alice
simply tests whether the correlation function between qubit 1,2 violate
Bell-inequality of two qubits or not. Then the security for the present
quantum secret sharing is guaranteed.

\subsection{Conclusion}

This investigation introduces a quantum secret sharing scheme using
four-qubit Smolin bound entangled state as private channel. The idea is very
similar to Ekert's 91 protocol about quantum key distribution \cite{Ekert91}%
. By testing the violation of the four-qubit correlation function to
two-setting Bell-inequality, the existence of eavesdropper can be detected.

Comparing with the QSS protocol via GHZ state in which a half of qubits must
be discarded because of incorrect direction for measurement, the efficiency
of the present scheme can reach $100\%$ if the quantum channel is secure.
Another QSS scheme with the help of bound entangled states is discussed in
Ref.\cite{Augusiak2}. However, its efficiency can only approach 50\% in
principle. On other hand, because each Bell state carries two bits hidden
classical information, for every copy of Smolin state Alice can send 2 bits
shared by Bob and Charlie.

But the generalization of this quantum secret sharing scheme to
multi-parties is not good as expected. In each Smolin state, the number of
qubits in hand of Alice is equal to the number of parties sharing secret
information. With the growing of the parties, the dimension of collective
measurement is increasing, while collective measurement on multi-qubit
system is more difficult to accomplish.

\subsection{Acknowledgments}

This work was financially supported by National Natural Science Foundation
of China under Grant. No. 10404007.

\end{document}